\documentclass[3p,times,twocolumn]{elsarticle}

\usepackage{ecrc}
\usepackage{xspace}

\newcommand{\pt}{$p_{\rm T}$\xspace}

\volume{00}

\firstpage{1}

\journalname{Nuclear and Particle Physics Proceedings}

\runauth{}

\jid{nppp}

\jnltitlelogo{Nuclear and Particle Physics Proceedings}

\usepackage{amssymb}

\usepackage[figuresright]{rotating}

\begin{document}

\begin{frontmatter}



\dochead{}

\title{Leading Hadron Production in $d$+Au and $^3$He+Au collisions\\ in the PHENIX experiment}


\author{Takao Sakaguchi, for the PHENIX collaboration}

\address{Physics Department, Brookhaven National Laboratory, Upton, NY 11973, USA}

\begin{abstract}
Neutral pions have been measured in $^3$He+Au collisions at
$\sqrt{s_{_{NN}}}$=200\,GeV up to 20\,GeV/$c$ in the RHIC Year-2014 run.
The nuclear modification factor $R_{\rm AA}$ was measured and compared
with that from $d$+Au collisions.  The integrated $R_{\rm AA}$ as a function
of $N_{\rm part}$ was calculated for $d$+Au, $^3$He+Au and Au+Au collisions
at $\sqrt{s_{_{NN}}}$=200\,GeV, and found to converge for
$N_{\rm part}>$12, while a clear system ordering
$R_{\rm dAu}$$>$$R_{\rm HeAu}$$>$$R_{\rm AuAu}$ was observed for
$N_{\rm part}<$12.
The fractional momentum loss for the most central $^3$He+Au collisions
was also estimated.
\end{abstract}


\end{frontmatter}


\section{Introduction}
The small collision systems such as $p/d$+A collisions have been
considered a good laboratory to quantify cold nuclear matter effects,
a necessary baseline for understanding the effects of the hot and
dense medium produced in A+A collisions. The observation of the
ridge-like structure in the long-range azimuthal correlations in $p$+Pb
collisions at $\sqrt{s_{_{NN}}}$=5.02\,TeV at the
LHC~\cite{Aad:2012gla, Aad:2014lta}, however, called into question
the view of such systems as consisting merely of cold nuclear matter.
The study at the LHC was followed by
the PHENIX experiment at RHIC, and a finite $v_2$ of hadrons in
0--5\,\% central $d$+Au collisions using both the two-particle angular
correlation method and the event-plane method were
shown~\cite{Adare:2013piz, Adare:2014keg}.
These observations led the community to explore any phenomena
found in A+A collisions, in $p/d$+A collisions.

The energy loss of hard scattered partons produced in the initial
stage of the collisions, so-called jet quenching, has been one of
the key observations
in confirming the production of the QGP. The first evidence of this
phenomena was found in the yield suppression of high transverse
momentum (\pt) hadrons,
the fragments of the hard scattered partons. The measurement of high
\pt identified hadrons have been improved over the last decade, and
reached to the level that a precise quantitative comparison of the
data and theoretical models became realized~\cite{Adare:2008cg, Adare:2012wg}.
A recent study also found that the energy loss scales with the
particle multiplicity ($dN/d\eta$)~\cite{Adare:2015cua}. Obviously,
the high \pt hadrons
will be a powerful tool to investigate the parton degree of freedom
in the small systems like $p/d$+A collisions as well.
In addition, a systematic study of the high \pt hadron spectra from
small to large collision systems will be able to explore the
onset of QGP as a function of the collision systems.
In this paper, we show the new measurement of the high \pt $\pi^0$ in
$^3$He+Au collisions and compared with the ones from $d$+Au and Au+Au
collisions and discuss the systematics of the yield suppression and
enhancement of the high \pt hadrons.

\section{Detector and Dataset}
\begin{figure}[ht]
\centering
\includegraphics[width=1.0\linewidth]{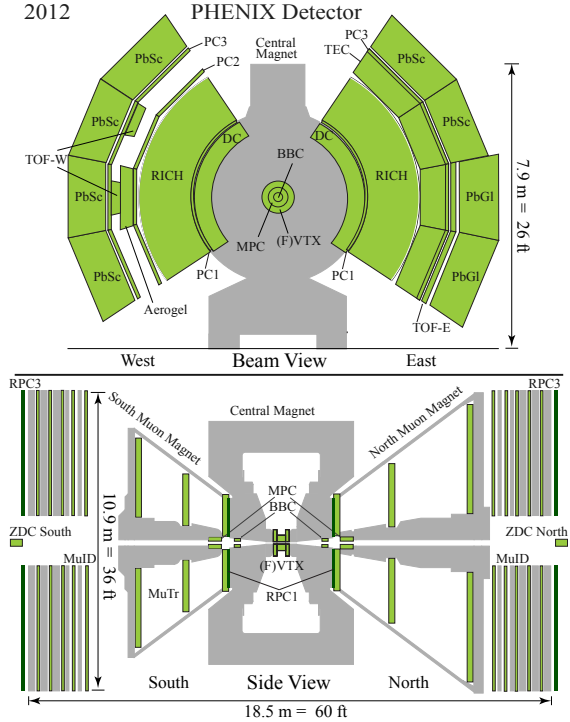}
\caption{PHENIX detector setup in the RHIC Year-2014 run.}
\label{fig:detector}
\end{figure}

PHENIX recorded an integrated luminosity of 25\,nb$^{-1}$ in
$^3$He+Au collisions at $\sqrt{s_{_{NN}}}$=200\,GeV in the RHIC
Year-2014 run. The detector setup was the same as the one in the
RHIC Year-2012 run as shown in Fig.~\ref{fig:detector}.
The detailed description of the PHENIX
detector system can be found elsewhere~\cite{Adcox:2003zm}.
The $\pi^0$ was reconstructed via $\pi^0\rightarrow\gamma\gamma$,
by primarily using a lead-scintillator sandwich type electromagnetic
calorimeter (PbSc EMCal). The threshold of cluster energy is set
to 0.2\,GeV and the photon clusters were selected using a shower
shape cut. Then, an energy asymmetry cut of
$\alpha=|E_1-E_2|/(E_1+E_2)<0.8$ was applied on selecting pairs
of photons from $\pi^0$ decay. The efficiency and acceptance of
the $\pi^0$ were estimated using a GEANT-based detector simulation
software.

Two types of the trigger selections were used to trigger events;
one is the coincidence of the signals from the two Beam-Beam
counters (BBC) located at 3.1$<|\eta|<$3.9 covering the full
azimuth (minimum bias trigger), and the other is the
coincidence of the minimum bias trigger and a high energy tower
hit in the EMCal (ERT trigger).
The events we used in this analysis were $2\times10^9$ minimum bias
triggered events, and $4.5\times10^{10}$ minimum-bias-equivalent
ERT events, totaling in the integrated luminosity of 22\,nb$^{-1}$.
The minimum bias trigger is not 100\,\% efficient to the inelastic
collisions because of the
limited acceptance and efficiency of the BBC. This inefficiency
increases as the collision system becomes smaller. They were
already studied in $d$+Au collisions by comparing the BBC charge
with a Glauber Monte Carlo simulation folded with a negative
binomial distribution~\cite{Adare:2013nff}. From this comparison,
we determined that the trigger efficiency is 88\,\% for $d$+Au
collisions. We followed the same method, and determined the
trigger efficiency of $^3$He+Au collisions also as 88\,\%.
In case of 200\,GeV Au+Au collisions, the efficiency was 94\,\%.
When dividing the events into centralities, an additional bias factor
plays a role. The bias is originated from the anti-correlation of
the available energies for producing particles in midrapidity
where the EMCal is installed, and the backward rapidity where the
BBC sits. We estimated the bias factors for the $^3$He+Au
collisions, also by following the method used for
$d$+Au collisions, and determined as 0.95, 1.02, 1.02, 1.03, and
0.89 for 0--20, 20--40, 40--60, 60--88, and 0--100\,\% $^3$He+Au
collisions, respectively~\cite{Adare:2013nff}.
We divided uncorrected yields by these factors.

\section{Results}
\begin{figure}[ht]
\includegraphics[width=1.0\linewidth]{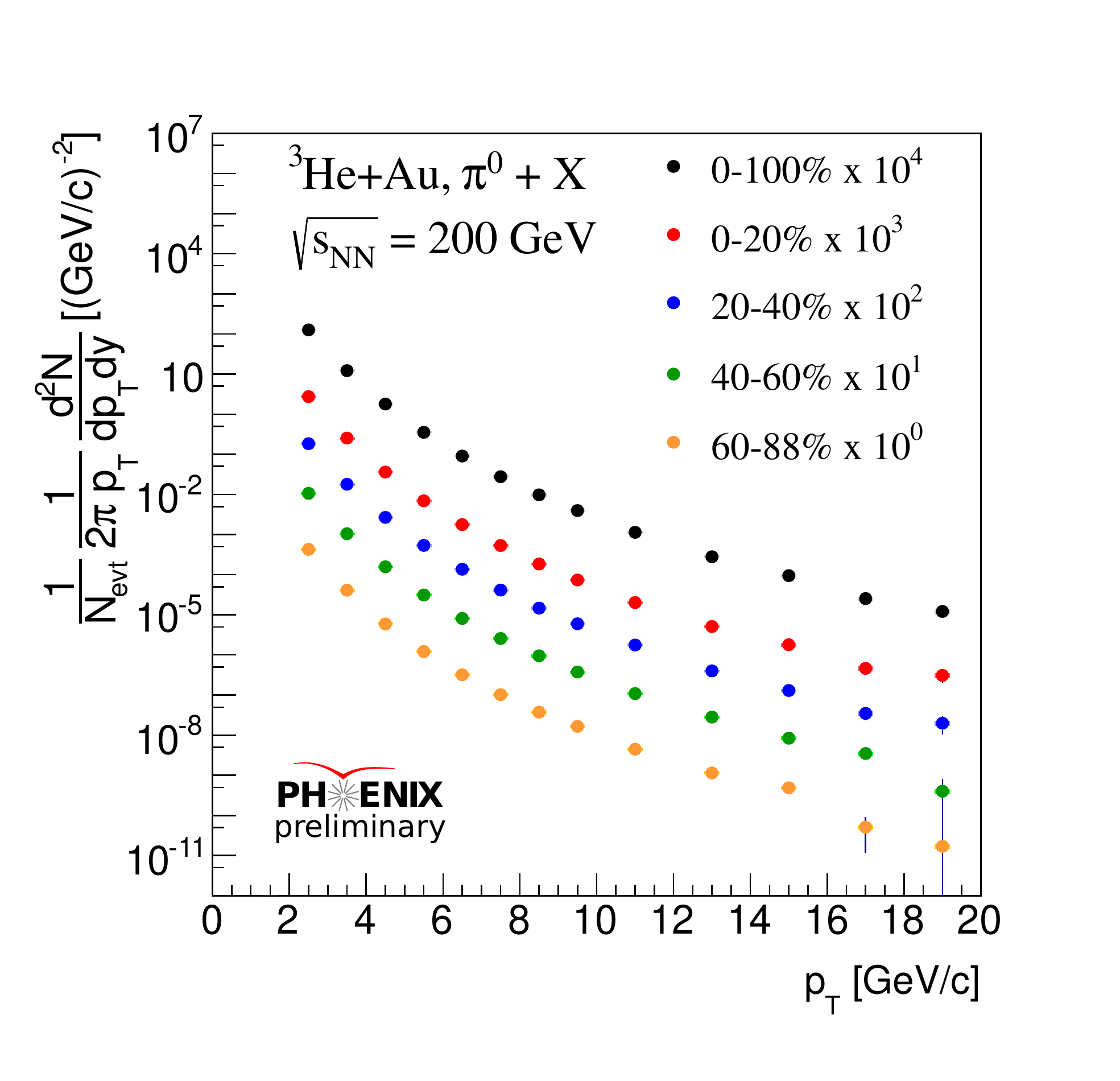}
\vspace{-7mm}
\caption{Invariant \pt spectra of $\pi^0$ in $^3$He+Au collisions
at $\sqrt{s_{_{NN}}}$=200\,GeV.}
\label{HeAuSpectra}
\end{figure}

\begin{figure}[ht]
\includegraphics[width=1.0\linewidth]{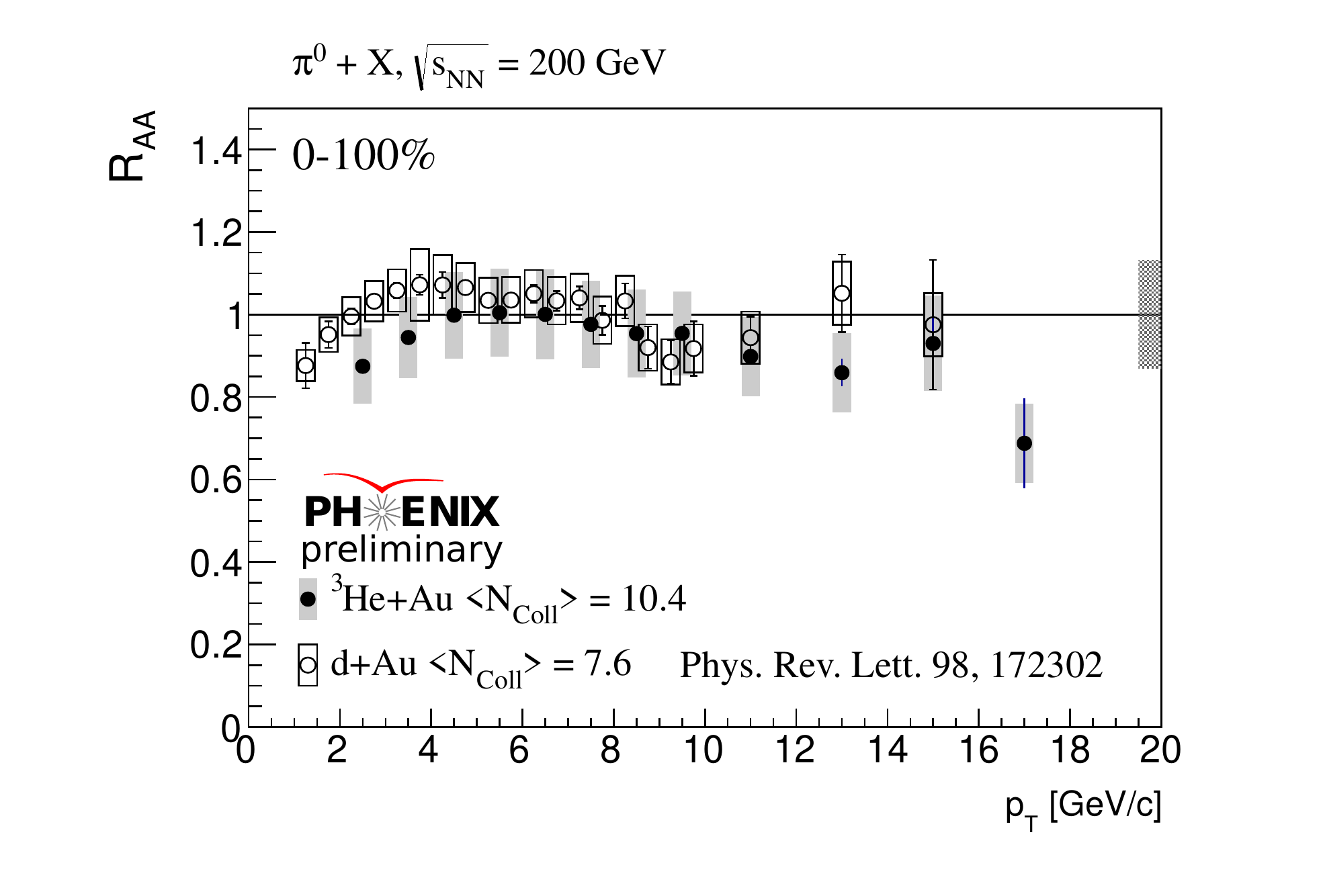}
\vspace{-7mm}
\caption{$R_{\rm AA}$ for 0--100\,\% $^3$He+Au collisions with the one from $d$+Au collisions.}
\label{HeAuRAA0_100}
\end{figure}

\begin{figure}[ht]
\includegraphics[width=1.0\linewidth]{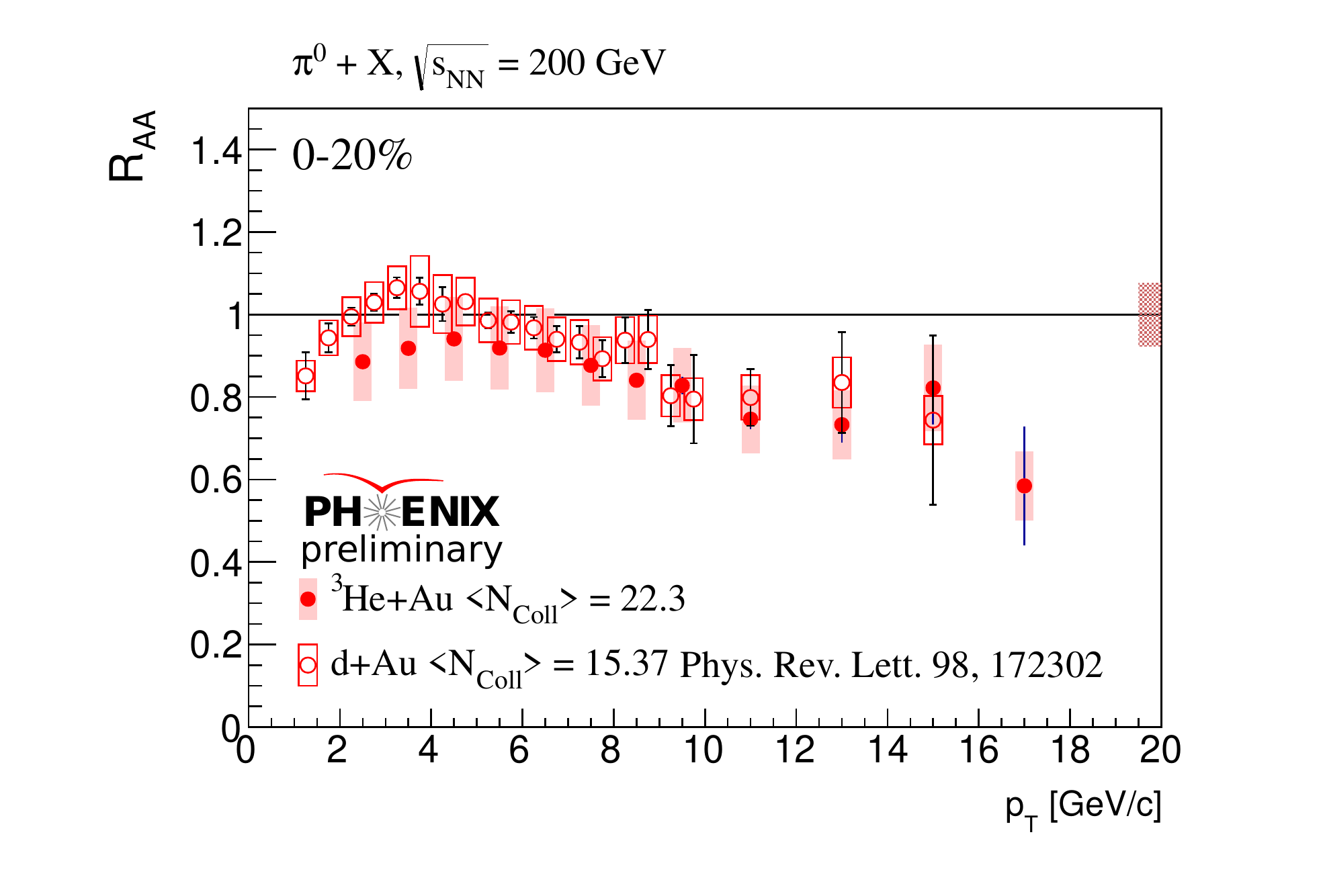}
\vspace{-7mm}
\caption{$R_{\rm AA}$ for 0--20\,\% $^3$He+Au collisions with the one from $d$+Au collisions.}
\label{HeAuRAA0_20}
\end{figure}

\begin{figure}[ht]
\includegraphics[width=1.0\linewidth]{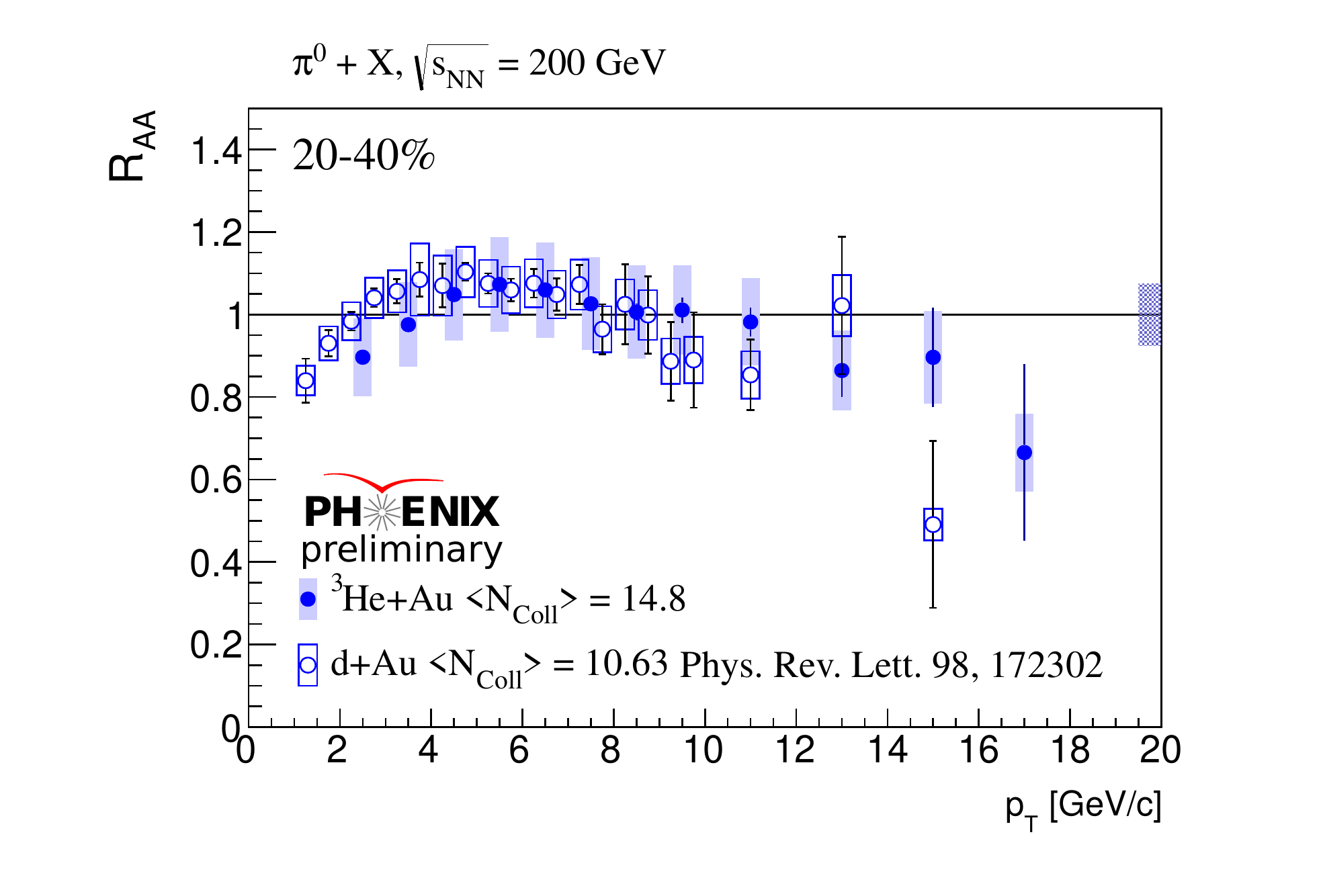}
\vspace{-7mm}
\caption{$R_{\rm AA}$ for 20--40\,\% $^3$He+Au collisions with the one from $d$+Au collisions.}
\label{HeAuRAA20_40}
\end{figure}

\begin{figure}[ht]
\includegraphics[width=1.0\linewidth]{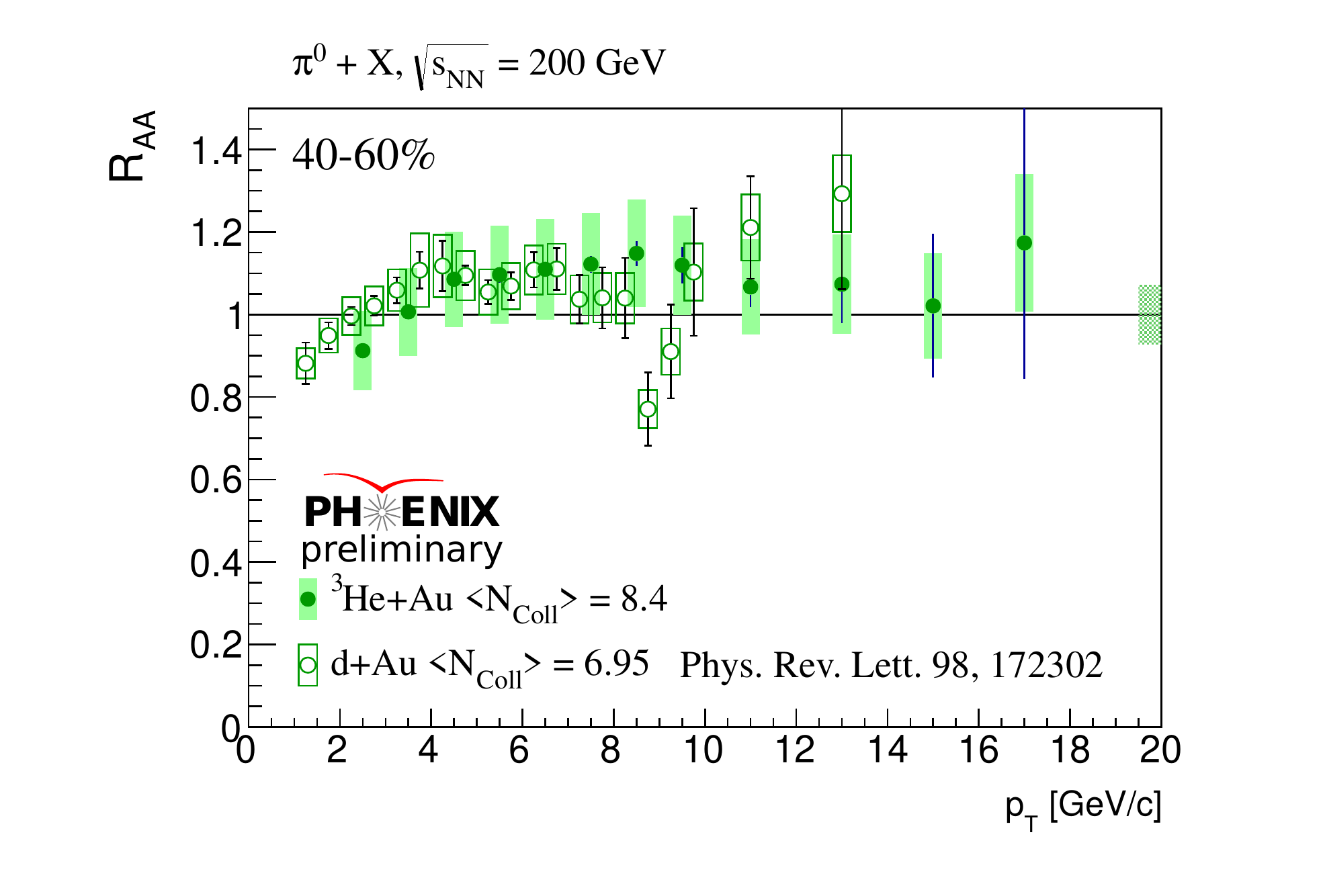}
\vspace{-7mm}
\caption{$R_{\rm AA}$ for 40--60\,\% $^3$He+Au collisions with the one from $d$+Au collisions.}
\label{HeAuRAA40_60}
\end{figure}

\begin{figure}[ht]
\includegraphics[width=1.0\linewidth]{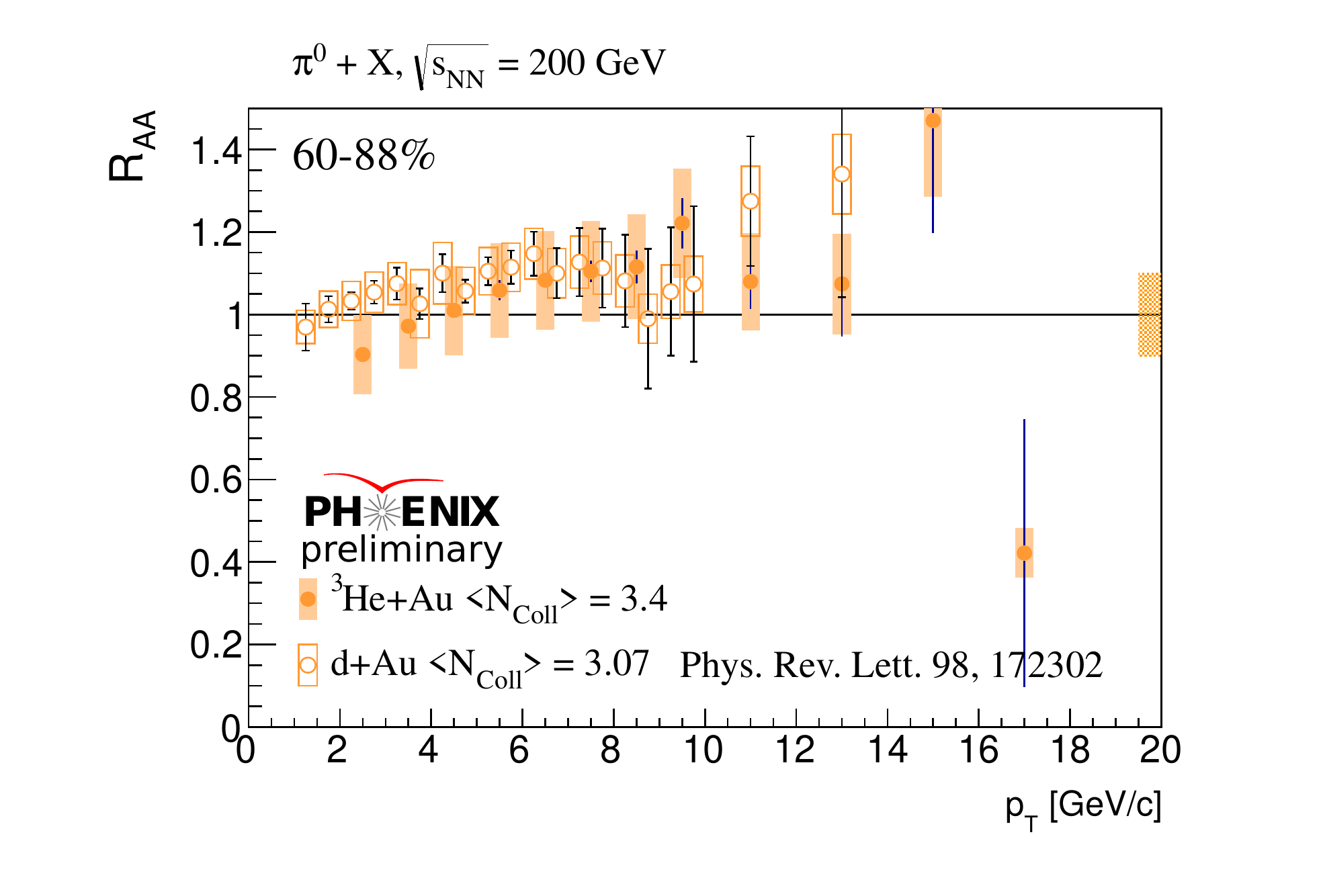}
\vspace{-7mm}
\caption{$R_{\rm AA}$ for 60--88\,\% $^3$He+Au collisions with the one from $d$+Au collisions.}
\label{HeAuRAA60_88}
\end{figure}

Fig.~\ref{HeAuSpectra} shows the invariant spectra of the $\pi^0$ as
a function of \pt in $^3$He+Au collisions at $\sqrt{s_{_{NN}}}$=200\,GeV.
The spectra from the minimum bias triggered events and the ERT
triggered events were connected at \pt=6\,GeV/$c$. The spectra were
successfully measured in the four centrality classes
(0--20, 20--40, 40--60, and 60-88\,\%) as well as the minimum bias events,
up to \pt=20\,GeV/$c$. The spectra are in a power-law shape suggesting
that the dominant source is the initial hard scattering.
In order to quantify the spectra change with respect to the one from
the primordial production ($p+p$ collisions), we computed the nuclear
modification factor ($R_{\rm AA}$) for all the five event classes.
We used the $\pi^0$ spectra from the $p+p$ collisions measured in the
RHIC Year-2005 run as the baseline~\cite{Adare:2007dg}.
Fig.~\ref{HeAuRAA0_100} shows the $R_{\rm AA}$ for the minimum bias
$^3$He+Au collisions.
The $R_{\rm AA}$ for the minimum bias $d$+Au collisions measured in
RHIC Year-2003 run are also shown for comparison~\cite{Adler:2006wg}.
In both systems, the $R_{\rm AA}$'s are consistent with unity within
quoted uncertainties. In Figs.~\ref{HeAuRAA0_20}, \ref{HeAuRAA20_40},
~\ref{HeAuRAA40_60}, and ~\ref{HeAuRAA60_88}, $R_{\rm AA}$'s for 0--20,
20--40, 40--60, 60--88\,\% $^3$He+Au collisions are shown, again together
with the ones in $d$+Au collisions, respectively.
The magnitude and the \pt dependence of the $R_{\rm AA}$ in the both
systems are remarkably similar within quoted uncertainties for all the
centrality classes. However, a small systematic difference is seen,
especially at the low \pt region (\pt$<$5\,GeV/$c$).

\begin{figure}[ht]
\includegraphics[width=1.0\linewidth]{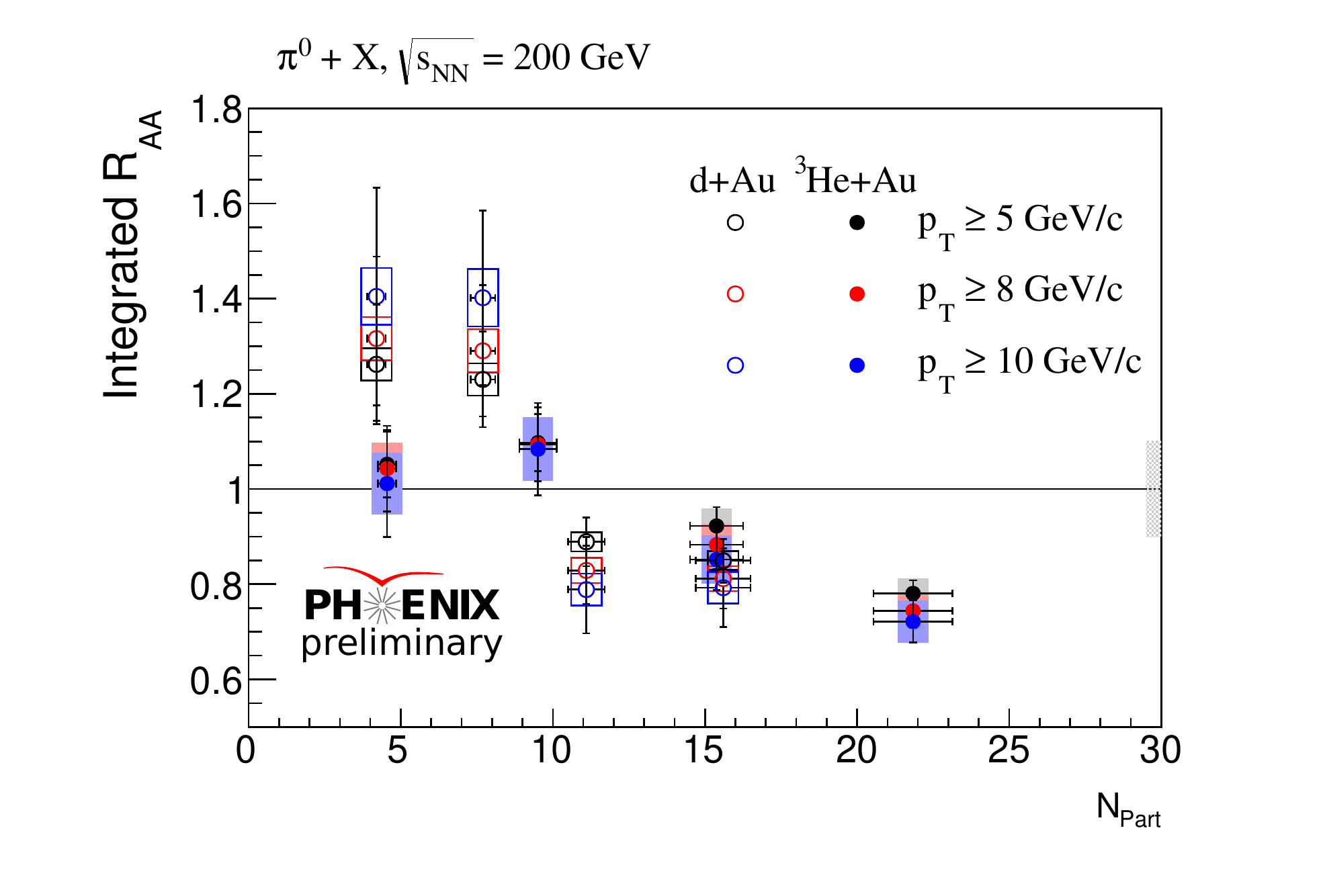}
\vspace{-7mm}
\caption{Integrated $R_{\rm AA}$ for $d$+Au and $^3$He+Au collisions.}
\label{IntegratedRAA}
\end{figure}

In order to systematically compare the enhancement/suppression of the
invariant yields in two systems, we plotted the integrated $R_{\rm AA}$
for $^3$He+Au and $d$+Au collisions as a function of number of
participant nucleons ($N_{\rm part}$) as shown in Fig.~\ref{IntegratedRAA}.

The integrated $R_{\rm AA}$ for three \pt ranges, namely
\pt$>$5\,GeV/$c$, $>$8\,GeV/$c$, and $>$10\,GeV/$c$, agree each other
for all the points, suggesting that the $R_{\rm AA}$ is flat over \pt
within the quoted uncertainties. At the higher $N_{\rm part}$, the
$d$+Au and $^3$He+Au collisions show the very similar $N_{\rm part}$
dependence, while at the lower $N_{\rm part}$, a clear distinction
between two collision systems are seen; the ones in $d$+Au collisions
show larger enhancement, suggesting a larger Cronin effect or less
suppression
(energy loss). We can study the trend even more systematically by
comparing those with the $R_{\rm AA}$'s from the peripheral Au+Au
collisions, i.e., 60--70, 70--80, and 80--93\,\% centrality.
In the previous publication, we measured the $R_{\rm AA}$ for
\pt$>$5\,GeV/$c$ as $\sim$0.78, $\sim$ 0.87, and $\sim$0.84,
for $N_{\rm part}$ of 26.7$\pm$3.7, 13.7$\pm$2.5, and 5.6$\pm$0.8,
respectively~\cite{Adare:2012wg}. From these numbers, we found
that the $R_{\rm AA}$'s from the three collisions systems converge
for $N_{\rm part}>$12, while a system ordering of
$R_{\rm dAu}$$>$$R_{\rm HeAu}$$>$$R_{\rm AuAu}$ is observed for
$N_{\rm part}<$12. The uncertainty needs to be improved in order to
make a concrete statement, but this finding suggests that a similar
medium, possibly a hot dense medium, may be created for the system
for $N_{\rm part}>$12.
Lastly, by comparing with the detail study of the fractional
momentum loss ($\delta$\pt/\pt) in Au+Au collisions carried out
in the previous publication, we found the the most central
$^3$He+Au collisions exhibit $\delta$\pt/\pt$\sim$0.03~\cite{Adare:2015cua}.

\section{Summary}
Identified $\pi^0$ has been measured in $^3$He+Au collisions at
$\sqrt{s_{_{NN}}}$=200\,GeV in the RHIC Year-2014 up to 20\,GeV/$c$
in four centrality classes as well as in the minimum bias events. The 
$R_{\rm AA}$ was computed and compared with those from $d$+Au collisions,
and found that they are consistent with quoted uncertainties, however, small
difference was seen between two systems, especially at the low \pt.
The integrated $R_{\rm AA}$ from $d$+Au, $^3$He+Au and Au+Au collisions
at $\sqrt{s_{_{NN}}}$=200\,GeV were found to converge for $N_{\rm part}>$12,
while a clear system ordering $R_{\rm dAu}$$>$$R_{\rm HeAu}$$>$$R_{\rm AuAu}$
was observed for $N_{\rm part}<$12, suggesting a similar medium
may be created for the system for $N_{\rm part}>$12.
We also found the most central $^3$He+Au collisions exhibited
$\delta$\pt/\pt$\sim$0.03.
The measurement in $p$+Au collisions at
RHIC will help completing this systematic study and could point the onset
of QGP in terms of the collision system size.



\bibliographystyle{elsarticle-num}
\bibliography{Sakaguchi_T}







\end{document}